\documentstyle[prl,aps,epsf]{revtex}
\begin{document}
\draft
\title{Direct current generation due to harmonic mixing:  From bulk
semiconductors to semiconductor superlattices}

\author{Kirill N.  Alekseev\cite{kirensky}\thanks{E-mail:
Kirill.Alekseev@oulu.fi}}
\address{Department of Physical Sciences,
Division of Theoretical Physics, Box 3000, University of Oulu
FIN-90014,
Finland}

\author{Feodor V.  Kusmartsev\cite{landau}\thanks{E-mail:
F.Kusmartsev@lboro.ac.uk}}

\address{ Department of Physics, School of
Mathematics and Physics, Loughborough University,
Leicestershire LE11 3TU,
United Kingdom}
\maketitle
\begin{abstract}
We discuss an effect of dc current and dc voltage (stopping bias) generation
in a semiconductor superlattice subjected by an ac electric field and its
phase-shifted $n$-th harmonic. In the low field limit, we find a simple
dependence of dc voltage on a strength, frequency, and relative phase of mixing
harmonics for an arbitrary even value of $n$.
\par
We show that the generated dc voltage has a maximum when a frequency of ac field is of the order of a scattering constant of electrons in a
superlattice. This means that for typical semiconductor superlattices at room
temperature operating in the  THz frequency domain the effect is really
observable. Indeed, such conditions are most used in the modern experiments
on THz harmonics generation in superlattices.
\par
We also made a comparison of a recent paper describing an effect of a directed
current  generation in a semiconductor superlattice subjected by ac field and
its second harmonic ($n=2$) [K. Seeger, Appl. Phys. Lett. {\bf 76}, 82 (2000)]
with  our earlier findings describing the same effect [K. Alekseev {\it et al.},
Europhys. Lett. {\bf 47}, 595 (1999); cond-mat/9903092]. In particular, here we
found that the maximum value of dc current is associated with the THz domain
that also follows from Seeger's calculations. This maximum value has already
been highlighted in our previous work.
\par
For the mixing of an ac field and its $n$-th harmonic with $n\geq 4$, we found
that additionally to the phase-shift controlling of the dc current, there is a
frequency control. This frequency controlling of the dc current direction is
absent in the case of $n=2$. The found effect is that, both the dc current
suppression and the dc current reversals exist for some particular values of ac
field frequency. For typical semiconductor superlattices such an interesting
behavior of the dc current should be observable also in the THz domain.
\par
Finally, we briefly review the history of the problem of the dc current
generation at mixing of  harmonics associated with arbitrary coherent
electromagnetic waves in semiconductors and semiconductor microstructures.
\end{abstract}

\section{Introduction}
It is natural that a coherent mixing of the waves with commensurate frequencies
(like, for example, having different harmonics) in a nonlinear medium can
result in a product which has a zero frequency or a static (dc) electromagnetic
field. If such a nonlinear interference phenomenon will happen in
semiconductors or semiconductor devices, like a semiconductor superlattice
(SSL), then the static electric field may result into a dc current or a dc
voltage generation. Such dc current (DC), which may be created in stationary
or nonstationary regimes, depends on amplitudes, frequencies, and relative
phases of mixing waves.
\par
It was Rosenblat \cite{rosenblat} who first at the end of 40th
suggested that the dc component of electric or magnetic
field may arise in pure ac-driven electrical circuits with nonlinear and
symmetric characteristics.  A similar idea has been proposed
by Skov and Pearlstein in the middle of 60th \cite{skov}.
Soon after these theoretical predictions, the DC due to mixing of ac
electric field and its second harmonic has been observed for the warm electrons
in  germanium by Pozhela and Karlin \cite{pozhela}, and, independently, by
Schneider and Seeger \cite{schneider}.
\par
Our paper is related to the same  issue of the DC generation and
consists of two parts. The first part has been stimulated by recent
paper \cite{seeger00} and contains original results on the DC generation
due to harmonic mixing in semiconductor superlattices, as well as a comparison
of the Seeger results \cite{seeger00} with the results obtained earlier in
our previous paper \cite{alekseev99}.
The second part is devoted to a short review of papers devoted to the
harmonic mixing induced DC in bulk semiconductors and SSLs. Here, a special
attention is paid to a consideration of different mechanisms of nonlinearity
in semiconductors leading to their non-ohmic behavior.

\section{The direct current due to a mixing of arbitrary harmonics in
a semiconductor superlattice}

An appearance of DC in SSL \cite{esaki70} under simultaneous action of ac
electric field and its phase-shifted second harmonic has been recently
shown by Seeger in the theoretical article \cite{seeger00}. He attributed this
DC effect to an ``Esaki-Tsu non-ohmicity of SSL for the case where it Bloch
oscillates'' and pointed out that a DC direction  depends on the phase-shift
\cite{seeger00}.
In the present paper we describe a DC arising in SSL due to mixing of
ac field and its phase-shifted $n$-th harmonic,  where $n$ is arbitrary
even number, like, $n=2,4,6,8,10$. Our main findings are following.
\par
First, we show that for a weak field strength, the DC is maximal only
when the mixing arises for waves having the THz frequencies. Thus, the effect
is relevant to a strong current interest to THz-field harmonics generation
in SSLs \cite{ghosh99,winnerl00}.
\par
Second, we show that for a mixing of ac field and its $n$-th harmonic with
$n\geq 4$ in typical SSLs there appear reversals of direction of the DC
and the DC may vanish for some particular frequencies, which values also belong
to the THz frequency domain. This novel mechanism of the  frequency-control is
supplementary to the control of magnitude and direction of DC using a variation
of phase shift. Moreover, the controlling of the phase shift is a difficult
task due to a very short dephasing time. On the other hand the frequency
control is easy to handle. The conditions for DC vanishing are different from
the well-known conditions for a dynamic localization
in ac-driven superlattices \cite{ignatov75,dunlap86,holthaus}.
The effect found is important because of a strong interest to high-order
harmonic generation in SSLs \cite{feise,pronin}.
\par
Consider a mixing of the first and the  $\phi-$phase-shifted $n$-th
harmonics of the ac electric field
\begin{equation}
\label{E_hm}
E(t)=E_1\cos(\omega t)+E_n\cos(n\omega t+\phi).
\end{equation}
in a single miniband SSL having the tight-binding energy-quasimomentum
dispersion relation
\begin{equation}
\label{tight-binding}
\varepsilon(k)=\frac{\Delta}{2}\left[1-\cos(k a)\right],
\end{equation}
where $k$ is an electron wave vector along the axis of SSL with the spatial
period $a$ and with the miniband width $\Delta$. The current induced in SSL is
proportional to the electron velocity $v=\hbar^{-1}\partial\varepsilon/
\partial k$. Using a standard semiclassical theory of the wave mixing in SSLs
\cite{esaki71,romanov72,orlov}, we obtain the following expression for the DC
\begin{equation}
\label{j_dc}
j_{\rm dc}=j_0 \sum_{\mu_1,\mu_2=-\infty}^{+\infty}
\sum_{\nu=-\infty}^{+\infty}
\frac{(\mu_1+n\mu_2) x\cos(\nu\phi)+\sin(\nu\phi)}{1+(\mu_1+n\mu_2)^2
x^2}
J_{\mu_1}(\xi_1) J_{\mu_2}(\xi_n) J_{\mu_1-n\nu}(\xi_1)
J_{\mu_2+\nu}(\xi_n),
\end{equation}
where $x=\omega\tau$, $j_0=\frac{\hbar \sigma}{e \tau a}$ with
$\sigma$ being a static, ohmic conductivity along the SSL's axis,
$\tau$ is a characteristic scattering time,
$J_{\mu}(\xi)$ is the Bessel function, and $\xi_1=(e a E_1)/
(\hbar\omega)$, $\xi_n=(e a E_n)/(n\hbar\omega)$. The Eq. (\ref{j_dc}) is a
direct generalization of the corresponding expression for DC derived in Ref.
\cite{alekseev99} for the case $n=2$ and is an exact result obtained from
the Boltzmann equation with a single constant relaxation time. This equation is
also consistent with a corresponding Eq. (6) of the Seeger's paper
\cite{seeger00}, where, however, a summation over the phase variable has been
neglected. In the  limit  of a weak field $\xi_1,\xi_n\ll 1$, and
using $J_l(\xi)\approx(\xi/2)^l (1/l!)$, from Eq. (\ref{j_dc}) we obtain
\begin{equation}
\label{dc}
\xi_{dc}=-f_n(x)\xi_1^n\xi_n \cos\phi + O (\xi_1^l \xi_n^k),
\quad l+k=n+3.
\end{equation}
Here, following the Ref.\cite{belinicher} we have introduced the stopping bias
as $E_{dc}=j_{dc}/\sigma$ and the scaled dc voltage as $\xi_{dc}=(e a E_{dc})/
(\hbar\omega)$. The nontrivial prefactor $f_n(x)$ is nonvanishing for any even
value of $n$ and has the following forms
\begin{equation}
\label{f2}
f_2(x)={\frac {3}{2}}\,{\frac {{x}^{2}}{1+5\,{x}^{2}+4\,{x}^{4}}},
\end{equation}
\begin{equation}
\label{f4}
f_4(x)={\frac {5}{4}}\,{\frac{{x}^{4}\left (-1+5\,{x}^{2}\right
)}{1+30\,{x}
^{2}+273\,{x}^{4}+820\,{x}^{6}+576\,{x}^{8}}},
\end{equation}
\begin{equation}
\label{f6}
f_6(x)={\frac {21}{32}}\,{\frac{{x}^{6}\left (
1+84\,{x}^{4}-35\,{x}^{2}\right )}{1+91\,{x}^{2}+3003\,{x}^{4}
+44473\,{x}^{6}+296296\,{x}^{8}+773136\,{x}^{10}+518400\,{x}^{12}}},
\end{equation}
\begin{equation}
\label{f8}
f_8(x)=\frac {9}{32}\frac{1}{\Psi_8(x)}
\left (126\,{x}^{2}+3044\,{x}^{6}-1-1869\,{x}^{4}\right ){x}^{8},
\end{equation}
$$
\Psi_8(x)=1+204\,{x}^{2}+16422\,{x}^{4}+669188\,{x}^{6}+14739153\,{x}^{8}+
$$
\begin{equation}
\label{Psi8}
173721912\,{x}^{10}+1017067024\,{x}^{12}+2483133696\,{x}^{14}
+1625702400\,{x}^{16},
\end{equation}
To derive these formulae for $f_n(x)$, we have performed symbolic computations
using MAPLE with a summation in all indices in Eq. (\ref{j_dc}) from -12 up to
12. We should note that, as follows from Eq. (\ref{dc}), the dc voltage
depends only on the first degree of the $n$-th harmonic's field strength $E_n$,
as well as on the phase difference between the first and the $n$-th harmonic
via cosine term. An interesting physical meaning has the nontrivial prefactor
$f_n(x)$.
\par
We start the analysis of the obtained expression with the case of $n=2$ [Eq.
(\ref{f2})]. The function $f_2(x)$ is always positive and has a maximum at the
value $x_{max}\approx 0.71$ with $f_2(x_{max})\approx 0.17$ (see Fig. 1a).
Note that an expression  analogous to Eqs. (\ref{dc}),(\ref{f2}) has
been obtain in our earlier work \cite{alekseev99}; the same form can be also
derived from Eq. (7) of the Ref. \cite{seeger00}.
\par
For $n\geq 4$ the functions $f_n(x)$ demonstrate  more complex behavior.
So, the function $f_4(x)$ may be both positive and negative(see Fig. 1b),
has a zero value at $x_0=5^{-1/2} \approx 0.45$, has a  maximum at
$x_{max}\approx 1.07$ [$f_4(x_{max})\approx 2.96\times 10^{-3}$]
and a minimum at  $x_{min}\approx 0.29$ [$f_4(x_{min})\approx-8.58\times
10^{-4}$]. In the limit $x\rightarrow 0$ this function has a negative
derivative.
\par
The function $f_6(x)$ (see, Fig. 1c) has two zeros at the value $x_0^{(1)}
\approx 0.62$ and the value
$x_0^{(2)}\approx 0.18$. Its global maximum and global minimum are located at
the value $x_{max}^{(1)}\approx 1.27$
[$f_6(x_{max}^{(1)})\approx 2.27\times 10^{-5}$]
and at the value $x_{min}^{(1)}\approx 0.4$
[$f_6(x_{min}^{(1)})\approx -1.18 \times 10^{-5}$], respectively.
The second small maximum, which is almost invisible on
the scale of Fig. 1c, exists at the value $x_{max}^{(2)}\approx 0.14$ with
the function value $f_6(x_{max}^{(2)})\approx 4\times 10^{-7}$.
In the limit $x\rightarrow 0$ this function has a positive derivative.
\par
The function $f_8(x)$ (see, Fig. 1d) is vanishing in three points, i.e.
it has three zeros at the value $x_0^{(1)}\approx 0.73$,
the value $x_0^{(2)}\approx 0.26$ and at the value  $x_0^{(3)}\approx 0.1$.
Its global maximum and minimum are located at the value
$x_{max}^{(1)}\approx 1.4$ [$f_8(x_{max}^{(1)})\approx 9.58\times 10^{-8}$]
and at the value $x_{min}^{(1)}\approx 0.48$ [$f_8(x_{min}^{(1)})\approx
-6.5\times 10^{-8}$], respectively. The other smaller maxima and minima exist
at the points: $x_{max}^{(2)}\approx 0.2$ with $f_8(x_{max}^{(2)})\approx
6.6\times 10^{-9}$ and $x_{min}^{(2)}\approx 0.08$ with $f_8(x_{min}^{(2)})
\approx -4\times 10^{-11}$.
In the limit $x\rightarrow 0$ this function has a negative derivative.
\par
All functions, $f_n(x)$, with $n\geq 4$ have maximums and minimums for
$x\simeq 1$ with $\max |f_n(x)|\simeq 10^{-n+1}$
and show asymptotically universal behavior:
$\propto x^n$ for $x\ll 1$ and $\propto x^{-2}$ for $x\gg 1$.
For typical SSLs at room temperature the scattering time
$\tau$  is of the order of $10^{-13}$ sec \cite{winnerl97,winnerl00},
therefore, the condition $x=\omega\tau\approx 1$ corresponds to a
frequency of about several THz. Thus, the most effective generation of DC
arises in the THz domain of frequencies. This is an important conclusion
missed in the Ref. \cite{seeger00}. The condition, $\omega\tau\approx 1$,
at which a maximal DC arises, is just typical for
experiments on the THz-field harmonic generation \cite{ghosh99,winnerl00}.
\par
The fact that the functions $f_n(x)$ with $n\geq 4$ can change a
sign means that the multiple DC reversals are possible not only by a change
of the phase difference $\phi$, but also {\it by a variation of the ac
frequency itself at the fixed phase difference}.
The multiple DC reversals in ac-driven SSL is a new effect; earlier such
a behavior of the DC has been described only in the ratchet systems, i.e., for
a particle moving in a periodic {\it asymmetric} potential in an external
ac field \cite{jung,mateos}.
\par
We also should note that the effect of the harmonic-mixing-induced DC vanishing
for a particular value of $x$ is different from the well-known dynamic
localization of electrons in ac-driven SSL
\cite{ignatov75,dunlap86,holthaus}\footnote{
Recently, the theory of a dynamical localization in tight-binding lattices
was generalized to the case of a two-frequency driven field (see,  Ref.
\cite{karczmarek}). In a comparison with the case of a single-frequency field
\cite{dunlap86,holthaus}, the localization in the two-color field is more
robust and takes place for a more wide
range of the field amplitude and the frequencies \cite{karczmarek}. However,
the effect is also observable only for $\omega\tau\gg 1$.
}.
Indeed, we have a suppression of the dc current component
in the limit of a weak field $\xi_{1,n}\ll 1$ and for $\omega\tau\simeq 1$,
while there in the dynamic localization arises a whole current suppression
for a scaled field strength greater or  of the order of unity and for
$\omega\tau\gg 1$ \cite{ignatov75,dunlap86,holthaus,karczmarek}.

\section{Historical preview}

In different semiconductors the different types of nonlinearity (or their
combinations) can result in the DC at harmonic mixing. The most universal
nonlinearity is caused by a nonparobolicity of conduction band. This type of
nonlinearity is mainly typical for the semiconductors and semiconductor
microstructures with the relatively wide bands
and narrow gaps, as well as for an opposite case of the narrow band and the
wide gap materials \cite{romanov72}. The later class includes SSLs. Another
main type of nonlinearity is the carrier's heating mechanism, which is mainly
responsible for the DC at wave mixing the semiconductors with wide band and
wide gap \cite{pozhela,schneider}.
Which type of nonlinearity would be dominant for the effects
associated with the wave mixing in many semiconductors with competing
nonlinearities, strongly depends on the temperature and the frequencies of
applied ac fields \cite{belyantsev71a}.
In the bulk semiconductors the effect of the DC generation
has been mainly considered for the mixing of microwaves, while for the SSLs
the main interest is paid to the THz frequency domain (submillimeter
wavelengths). In this part of the work we briefly review both situations.

\subsection{Wave mixing in bulk semiconductors}

The DC at the mixing of microwaves having commensurate frequencies in such
semiconductors as n-Ge and n-Si has
been intensively studied both experimentally and theoretically by Vilnus
\cite{pozhela,banis66,banis70,dargis71,dargis72,bondarenka} and
Wien \cite{schneider,seeger70,mayr} groups.
The experiments were performed at 77 K or at room temperatures
and the majority of them dealt with the mixing of the fist and the second
harmonics of coherent microwaves, although the effect of mixing of even
higher harmonics (up to 6th) has been also considered in Ref.\cite{banis70}.
The main nonlinear mechanism responsible for a generation of the DC at wave
mixing in these semiconductors was found to be the electron's heating by
the field. It has been shown that a simple model based on  balance equations
and incorporating a parabolic band and the dependence
of the carrier's relaxation time on the field (or on their energy), can well
describe the effect in Ge and Si \cite{seeger70}. It was also suggested that
the effect is suitable for a determination of different carrier's relaxation
times \cite{seeger70,dargis71,dargis72}.
\par
Independently from this research the DC generation  in semiconductors
associated with a mixing of a coherent electromagnetic wave with its second
harmonic has been also described in theoretical papers
\cite{shmelev71,entin89,baskin} and named as ``coherent photovoltaic effect''.
The description based on a quantum kinetic equation for the model of free
electrons interacting with acoustical phonons was used in the paper
\cite{shmelev71}, while the paper \cite{entin89} dealt with
an approach of the classical Boltzmann equation  with  collisional
integrals incorporating a field absorption  by free classical electrons
or by impurity-band transitions. Finally, the paper \cite{baskin} suggested
that a mechanism of the DC generation at the wave mixing can be associated
with quantum corrections to  a nonlinear rf-conductivity.
We should stress that in all enumerated papers the band has been assumed to
be parabolic and the nonlinearity was caused by different forms of the
dependence of the carrier's relaxation time on the electron energy or on
the field strength.
\par
However, the experiments of  Patel {\it et al.} \cite{patel} have clearly
demonstrated that an optical waves mixing in several III-V type
semiconductors with narrow gaps, such as n-doped InAs, InSb and GaAs, can
not be explained without taking into account the miniband nonparabolicity
as a source of nonlinearity within the material\footnote{
Also note that nonparobolicity of electron dispersion law as possible
mechanism for the  strong third-order optical mixing has  earlier been
pointed out by Lax {\it et al.} \cite{lax}.
}
\cite{wolff,wynne,bierig}. In these narrow-gap semiconductors the conduction
band has a strong nonparabolicity due to the strong interaction with valence
band. It is conventionally described by a Kane two-band or a four band model,
which has a strong nonlinear dispersion law \cite{kane,bir}.
\par
Later on, several experiments on the mixing of mm-waves in the same materials
at liquid helium and 77K temperatures have been performed
\cite{belyantsev70,belyantsev71,kozlov72,belyantsev73,leonov}.
Their analysis indicate that for microwaves both nonlinearity mechanisms, due
to the non-quadratic dispersion law in the conduction band and to changes in
the collision time because of electron heating, are important
\cite{belyantsev71,belyantsev71a,genkin}.
The estimates presented in Refs. \cite{belyantsev71a,genkin} for such
semiconductors as n-InSb and n-GaAs demonstrate that for the high frequencies
($\gtrsim 10^{13}$ ${\rm s}^{-1}$) and moderate temperatures, the so-called
``dynamical nonlinearity'' (band nonparabolicity)
prevails, in a contrast to a case of  low frequencies and low temperatures,
where the heating nonlinearity  mainly predominates.
\par
Finally, an experiment on the third harmonic generation of a submillimetere
radiation ($\lambda_{\omega}=0.9$ mm, $\lambda_{3\omega}=0.3$ mm) in the
$n$-type indium antimonide has been described in the Ref. \cite{belyantsev81}.
There the main mechanism of a nonlinearity driving the wave mixing was
identified as being related to a nonparobolicity of the conduction band.
However, to the best of our knowledge, the possibility of the DC generation
due to a harmonic mixing of THz fields in such narrow gap bulk semiconductors
was recently discussed only in our paper \cite{alekseev99}.

\subsection{Wave mixing in semiconductor superlattices}

Another class of semiconductors, in which the band nonparabolicity should
be important, is associated with materials having narrow bands and wide gaps,
such as SiC and ZnS. Recently a new class of the systems with narrow bands
appear: Arrays of quantum dots or quantum wells. One of the most popular
and most studied system is a semiconductor superlattice (SSL) consisting of
a series parallel layers with quantum wells. In general the SSL \cite{esaki70}
may be viewed as an artificial crystal having narrow (mini-)bands and
relatively wide gaps. Nowadays these semiconductor microstructures attracts
much attention \cite{grahn-book}.
\par
Already in the first paper on SSLs, Esaki and Tsu suggested that
the miniband nonparabolicity is a main source of nonlinearity, which can
be used for an observation of Bloch oscillations and a negative
differential conductivity \cite{esaki70}. For low or moderate field strengths,
the nonlinear transport properties of a single miniband SSL could be well
described in the framework of a semiclassical Boltzmann equation with a
constant relaxation time \cite{esaki70,bass}.
The theoretical predictions obtained within such an approximation are
in a good agreement with the measurements of voltage-current characteristics
in SSLs for a temperature above 40 K \cite{grahn91,sibille}.
On the other hand, there is an important theoretical result by
Suris and Shchamkhalova \cite{suris}, which states that a collisional
integral in the quantum kinetic equation
is practically independent  on the field in the low field limit.
This and other theoretical results \cite{bryksin97,bryksin99}
give a support for an application of a constant relaxation time
approximation for low field and high temperatures.
\par
In 1971 Esaki and Tsu \cite{esaki71} and Romanov \cite{romanov72}
suggested to use the SSL as a new, artificial, nonlinear material
for electromagnetic wave mixing and new harmonics generation. They showed
that strong SSL's miniband nonparabolicity could be source of nonlinearity
for harmonics generation in analogy with the case of bulk semiconductors
studied in Refs. \cite{patel,wolff}. The theory of wave mixing in SSLs,
based on a solution of the Boltzmann equation with a constant relaxation time
for single miniband, has been developed in works
\cite{ignatov76,pavlovich76,orlov,romanov78} (see also \cite{bass}).
Our result on the DC generation due to  mixing of the first and the second
harmonic \cite{alekseev99}, as well as corresponding results by Seeger
\cite{seeger00}, could be immediately derived using the results of
previous paper by Orlov and Romanov \cite{orlov}.
\par
Earlier the DC due to the mixing of ac field and its second
harmonic in the tight-binding lattice has been described by Goychuk and
H\"{a}nggi \cite{goychuk98} using a theory of quantum ratchets. The so-called
ratchet system consists of a particle moving in a periodic {\it asymmetric}
potential and under the action of an external time-dependent force with a zero
average \cite{magnasco,ajdari}. The force can include a noise or can be pure
deterministic and spatially uniform (``rocking ratchets'') \cite{hanggi}. Both
types of ratchets demonstrate a rectification effect.
For a recent realization of the ratchet system in an antidot array of
SSL driven by a far-infrared field, see, Ref. \cite{lorke}.
Another mechanism for the DC generation due to a mixing of first and second
harmonics in a quantum tight-binding lattice has been
considered in the Ref. \cite{karczmarek}. Following \cite{karczmarek}
the DC can be generated if the intersite coupling constants of the lattice
are linearly dependent on a lattice site number.
\par
Among other recent theoretical results related to the DC arising at
the harmonic mixing in superlattices, we would like to mention a consideration
of SSL driven by a monochromatic ac field
\cite{alekseev98,alekseev98a,alekseev96,ignatov95,cao}.
There, in single miniband of SSL   a self-consistent field is
generated by electron motion \cite{epshtein78,romanov79,alekseev94}.
Within this model the transport properties of SSL
are effectively controlled by a mixing of an external, pure harmonic
driving field and the self-consistent field \cite{alekseev99,alekseev98}.
\par
For an ac field strength which is lower than a some critical value, a Fourier
spectrum of the self-consistent field contains only odd harmonics and,
therefore, the DC can not be generated. However, for an external ac field,
which is strong enough, the self-consistent  field is generated and its Fourier
spectrum contains even harmonics as well. The mixing  of  the even and the odd
harmonics results in the DC \cite{alekseev99}. For a relatively weak
scattering of electrons with impurities and phonons  the self-generated current
is chaotic \cite{alekseev98a,alekseev96,cao}, while for a strong scattering it
is quantized \cite{alekseev98,alekseev98a,ignatov95}. For further development
of the wave-mixing theory in a pure ac-driven SSL see recent work
\cite{romanov00}.
\par
Another interesting theoretical direction is an investigation of symmetries
and the symmetry breaking for a particle moving in  a spatially periodic
potential under an action of a time-periodic force
\cite{flach,yevtushenko00,goychuk00,yevtushenko00a}.
In particular, the direct current may arise at the harmonic-mixing even if an
underlying motion of the particle in a cosine-like spatial potential is chaotic
(dynamical chaos) \cite{flach,goychuk00}. However, in semiconductor
microstructures, as far as we aware, this interesting theoretical suggestion
has not been realized, yet.

\section{Conclusion}

In summary, the results of papers \cite{alekseev99,seeger00}, together
with a consideration based on the theory of quantum ratchets \cite{goychuk98},
demonstrate that in semiconductor superlattices the effect of a dc current
(voltage) generation due to a mixing waves associated with different
commensurate harmonics is now well established theoretically.
The source of the non-ohmicity in current-voltage characteristics
of the strongly-coupled SSLs may be associated with many factors. This may
primarily be related to a nonparabolic band energy-momentum relation.
This type of nonlinearity exists in any  semiconductor or
in any semiconductor device and does not depend on the fact if it is subjected
by an intense high-frequency electric field or not.
\par
We believe that this paper attracts attention of wide scientific community
to these old and at the same time novel effects, will stimulate
new experiments and a construction of new devices associated with a
rectification and a detection of the THz radiation. The experimental conditions
for an observation of the dc current effect are practically identical to those
fulfilled in a recent experiment on  a generation of harmonics of the THz
radiation in a semiconductor superlattice \cite{winnerl00}.
\par
We are thankful to  M.  Erementchouk, E.  Cannon, S.  Flach, I.
Goychuk, M. Saarela, P. Pietil\"{a}inen, O. Yevtushenko for discussions
and D. Campbell and P. H\"{a}nggi for stimulating interest.
This research was partially supported by Academy of Finland (grant 163358).

\vspace{1.5cm}
\epsfxsize=16cm
\hspace{0.5cm}
\epsfbox{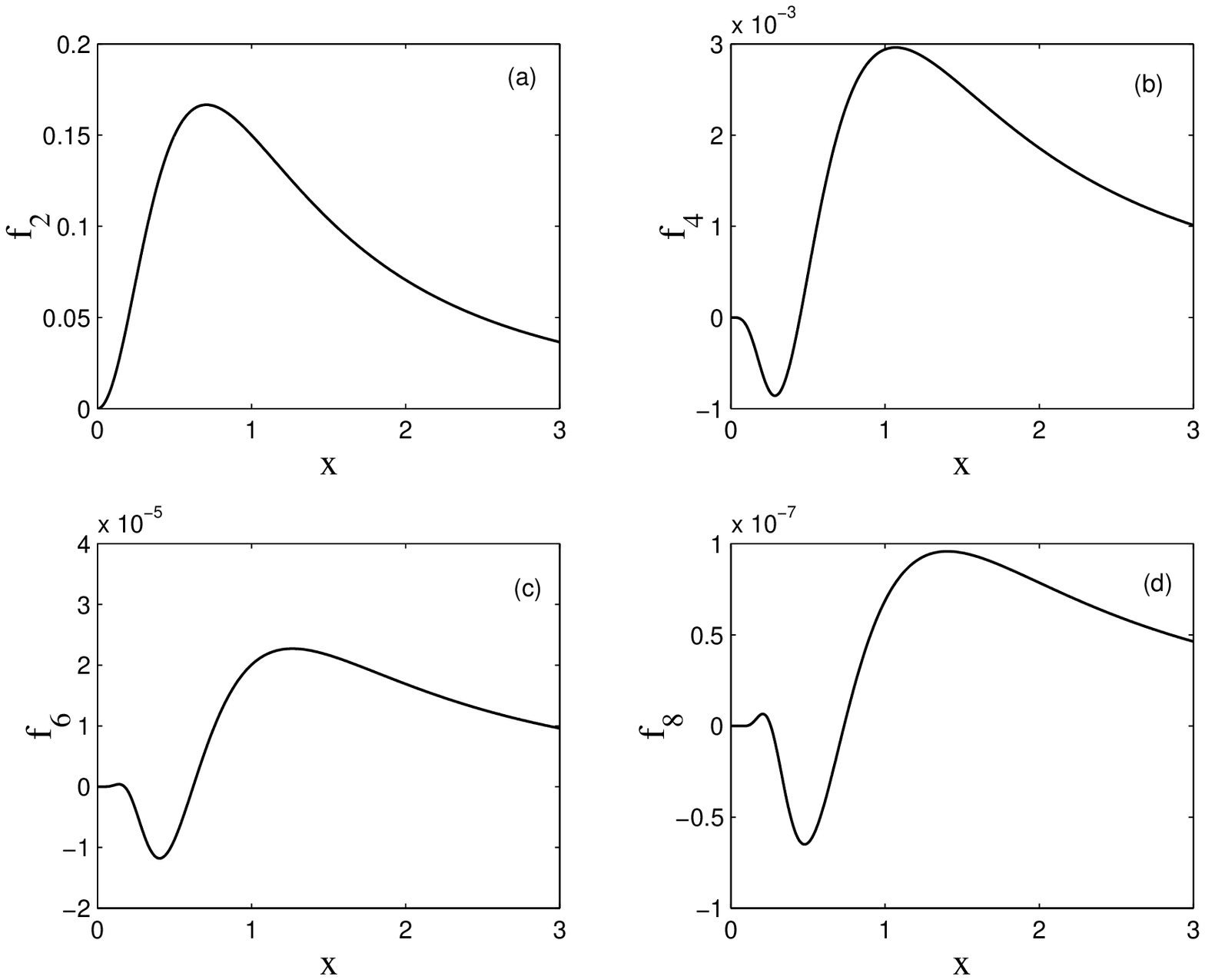}
\begin{figure}
\caption{The function $f_n(x)$ for $n=2$ (a), $n=4$ (b), $n=6$ (c),
and $n=8$ (d).}
\label{fig1}
\end{figure}

\end{document}